\pdfoutput=1
\documentclass{JINST}

\title{Scintillation efficiency of liquid argon in low energy neutron-argon scattering}
   
\author{W.~Creus$^a$\thanks{This work is based on the PhD Thesis of William Creus}, Y. Allkofer$^a$\thanks{Now at Credit Suisse AG, Zurich}, C. Amsler$^b$\thanks{Corresponding
author}, A. D. Ferella$^c$, J. Rochet$^b$,  L.~Scotto-Lavina$^d$, and M. Walter$^a$\\
\llap{$^a$}Physik-Institut der Universit\"at Z\"urich, CH-8057 Z\"urich, Switzerland\\
\llap{$^b$}Albert Einstein Center for Fundamental Physics,  
Laboratory for High Energy Physics, University of Bern, CH-3012 Bern, Switzerland\\
\llap{$^c$}INFN Laboratori Nazionali del Gran Sasso, LNGS, Assergi, Italy\\
\llap{$^d$} SUBATECH, Ecole des Mines de Nantes, CNRS/In2p3, Universit\'e de Nantes, Nantes, France

E-mail: \email{claude.amsler@cern.ch}}

\abstract{Experiments searching for weak interacting massive particles with noble gases such as liquid argon require very low detection thresholds for nuclear recoils. A determination of the scintillation efficiency is  crucial to quantify the response of the detector at low energy. We report the results obtained  with a small liquid argon cell using a monoenergetic neutron beam produced by a deuterium-deuterium fusion source. The light yield relative to electrons was measured  for six argon recoil energies between 11 and 120 keV at zero electric drift field.}

\keywords{Dark matter; WIMP; liquid argon; quenching factor; monoenergetic neutrons}

\begin{document}

\section{Introduction}
It is known from astronomical observations that most of the gravitational mass in the universe is made of dark energy and non-baryonic dark matter which does not couple to electromagnetic radiation \cite{Porter2011}. Dark matter has survived since the birth of the universe and hence must  be stable and weakly interacting. The most prominent candidates for dark matter are the Weakly Interacting Massive Particles (WIMPs) \cite{Jungmann1996}, in particular  the spin-1/2 neutralino  with  predicted mass  in the  GeV to   TeV range.  The neutralino would scatter on constituent quarks in nucleons, leading to nuclear recoils in the range of 1 -- 100 keV. Non-accelerator laboratory (``direct'') searches are all based on the detection of  nuclear recoils. The scattering cross section on nucleons is tiny, in the range 10$^{-5}$ to 10$^{-12}$ pb, comparable to that for  neutrino interaction. For low masses the sensitivity decreases due to the low recoil energy and the detection threshold, while for high masses the loss of sensitivity is due to the exponentially diminishing WIMP flux. 

The laboratory observation of dark matter  is one of the most pressing issues in Particle Physics. The neutralino is being searched for at the LHC  and direct searches for WIMPs are underway in non-accelerator underground experiments (for a review see \cite{DreeS$_2$014}). The most stringent upper limits for WIMPs are obtained for the spin independent WIMP-nucleon cross section. Upper limits on  the WIMP-nucleon cross section have been obtained by the XENON100 \cite{XENON} and LUX \cite{LUX} experiments using liquid xenon, the latter quoting the most stringent upper limit of 7.6 $\times$ 10$^{-10}$ pb for a WIMP mass of about 33 GeV. 

At the LHC the neutralino would manifest itself by leaving a large missing energy when scattering off quarks. These ``indirect'' searches at the LHC are more competitive than direct searches for low WIMP masses, providing the best upper limits of about 10$^{-3}$ pb below 3.5 GeV \cite{CMS$_2$012,CMS$_2$014}. 

Several direct searches were performed or initiated during the last ten years, employing different types of detectors such as solid state detectors or noble liquids (argon, neon or xenon) \cite{DreeS$_2$014} which can be used in large volume time projection chambers. Experiments using liquid argon are underway: DEAP \cite{DEAP} and MiniCLEAN \cite{MiniCLEAN} at SNOLAB, DarkSide \cite{DarkSide} at Gran Sasso, and ArDM \cite{ArDM} at Canfranc.

\section{Liquid argon for direct searches}
\label{sec:Fp}
\begin{figure}[htb]
\begin{center} 
\includegraphics[width=0.60\textwidth]{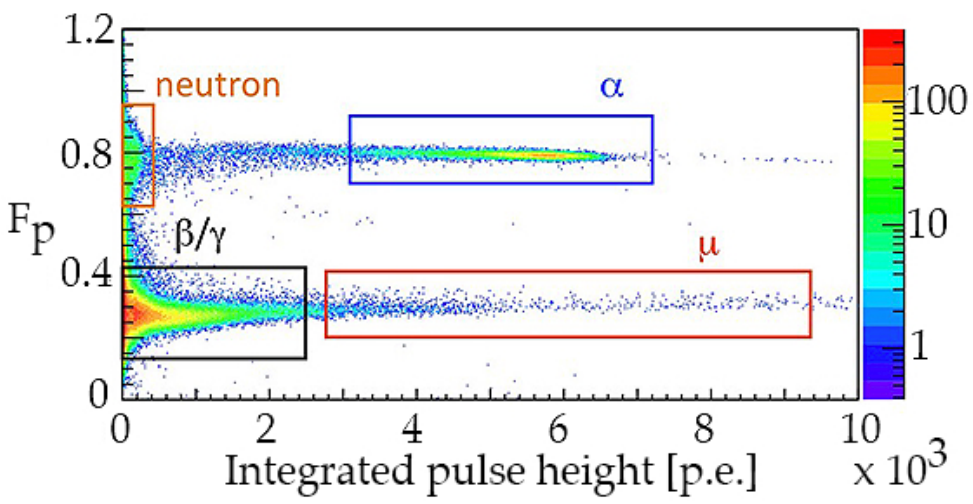}
\caption[]{Fraction $F_p$ of prompt to  total light for various projectiles.}
\label{Fp}
\end{center}
\end{figure}

In liquid argon (LAr) scintillation light is emitted in a narrow VUV band around 128  nm with two components of  different decay times,  $\tau_{1}\simeq 7$ ns and $\tau_{2}\simeq 1.6$ $\mu$s \cite{Hitachi}.  Heavily ionizing projectiles such as $\alpha$ particles or nuclear recoils contribute mostly to the fast decaying component, while the contribution from electrons and $\gamma$-rays to the slow component is larger.  In this work we define the prompt fraction $F_p$ as the fraction of prompt over  total integrated pulse height, the prompt light being chosen  as the integrated pulse height in the time interval between $-20$ and $+30$ ns of the time at which the light pulse reaches its maximum.  Figure \ref{Fp} shows the prompt fraction $F_p$ measured during the early stage of the experiment, when a $^{210}$Po $\alpha$-source (5.3 MeV) was installed \cite{Creus}. $F_p$  is typically  0.25 and 0.75 for electron recoils and nuclear recoils, respectively. The prompt fraction is therefore useful as a discriminant to identify nuclear recoils and to reject background \cite{Lippincott2008}. This property is the main advantage of argon over xenon (in addition to its lower price), but the disadvantage is the presence of the radioactive $\beta$-emitter $^{39}$Ar.

The scintillation light yield and its dependence on nuclear recoil energy is an essential input to the WIMP-nucleon cross section. The relative scintillation efficiency  $L_{eff}$  is defined as the ratio of the scintillation yield $L_{nr}$ from nuclear recoils to  that from electron recoils $ L_{er}$,
\begin{equation}
L_{eff} =\frac{L_{nr}(T_{nr})}{L_{er}(T_{er})},
\label{eq:Leff}
\end{equation}
where $T_{nr}$ and $T_{er}$  stand for the nuclear recoil and electron recoil energy, respectively. $ L_{er}$ is determined from photopeaks or Compton edges, using  $\gamma$-sources. The relative efficiency  
$L_{eff}$ being an energy dependent quantity, a standard reference for calibration is required. For the sake of convenience  we use in this work the 60 keV electronic photopeak from a  $^{241}$Am source.  As discussed below, we have measured the light yield  from various $\gamma$-sources  (including a 32 keV $^{83\mathrm{m}}$Kr source) and have established its linearity for recoil electrons above 30 keV. The results obtained here can therefore be compared directly with those obtained from earlier experiments  \cite{WARP,MicroCLEAN,SCENE} using e.g.  the $T_{er}$ = 122 keV photopeak from a $^{57}$Co source  \cite{MicroCLEAN}. 

The relative scintillation efficiency is measured by mimicking the WIMP-nucleus interaction with neutron beams, e.g.  by elastic scattering of monoenergetic neutrons under fixed scattering angles (see \cite{MicroCLEAN,SCENE} for argon and  \cite{GPlante} for xenon data). The measured light yield is then compared with a simulation and $L_{eff}$ is derived by applying iterative fits. This article reports  the determination of $L_{eff}$ in LAr at zero electric field using a monoenergetic neutron beam from deuterium-deuterium fusion. Details can be found in \cite{Creus}. The measurements were motivated by the envisaged DARWIN dark matter project \cite{DARWIN}. 

\subsection{Experimental setup}
\begin{figure}[htb]
\begin{center} 
\includegraphics[width=0.50\textwidth]{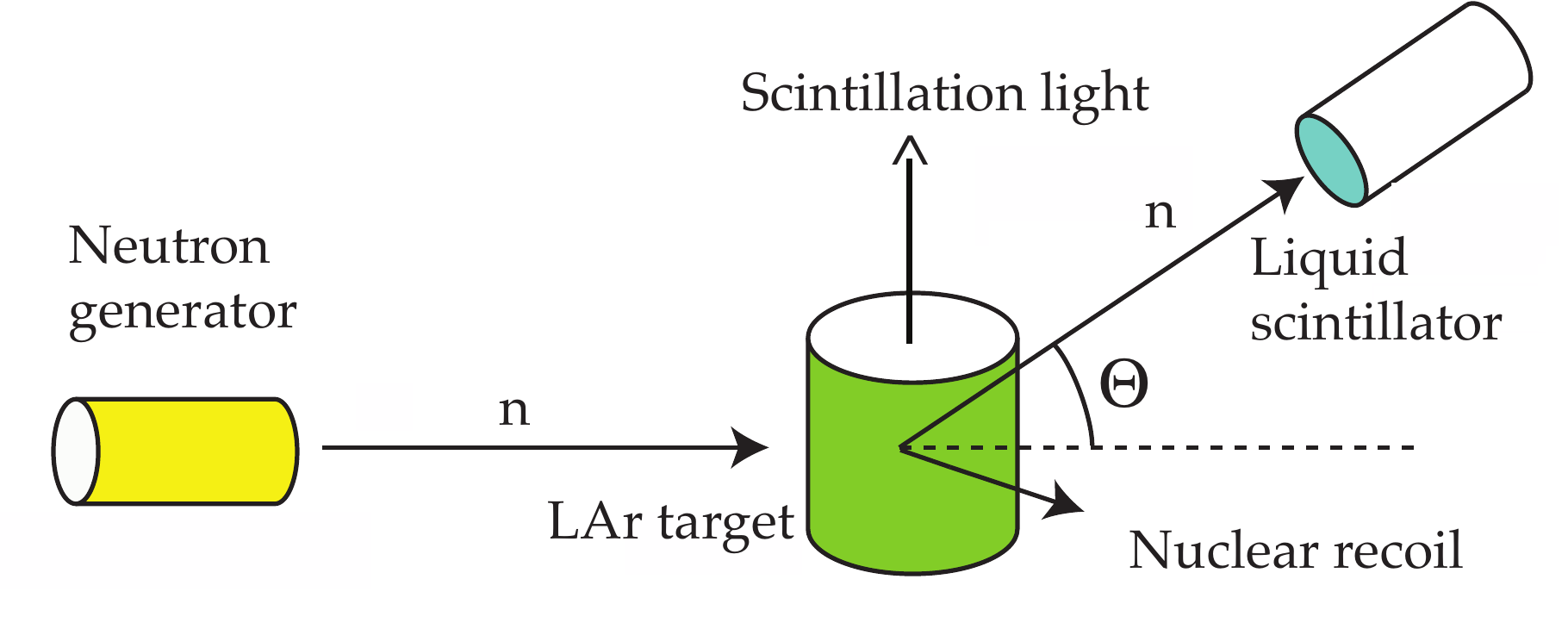}
\includegraphics[width=0.30\textwidth]{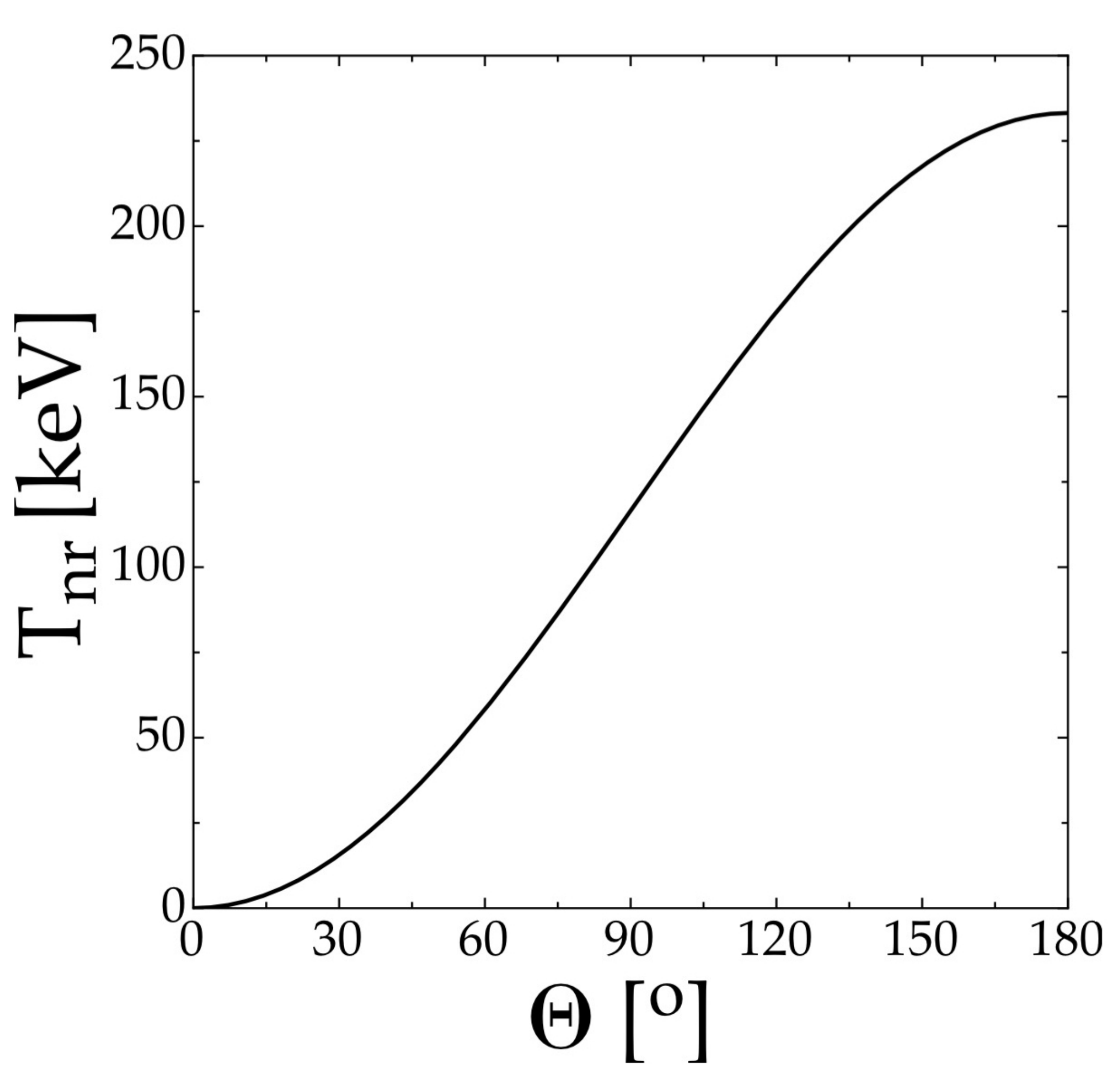}
\caption[]{Left: principle of the scattering experiment with monoenergetic neutrons scattered under the angle $\Theta$. Right: nuclear recoil energies in argon for incident neutron energies of 2.45 MeV.}
\label{Principle}
\end{center}
\end{figure}

A good way to calibrate the light output and to study the response of LAr to nuclear recoils is to scatter a beam of monoenergetic neutrons of energy $T_n$ off argon nuclei, and to measure the light yield as a function of scattering angle $\Theta$, from which the recoil energy $T_{nr}$ can be calculated according to the formula

\begin{equation}
T_{nr} \simeq \frac{2T_nA}{(1+A)^{2}}\left(1-\cos\Theta\right),
\end{equation}
where $\Theta$ is the scattered neutron angle in the laboratory and $A  \gg 1 $ is the atomic mass number. The method is illustrated in figure \ref{Principle} (left). 

\begin{figure}[htb]
\begin{center} 
\includegraphics[width=0.55\textwidth]{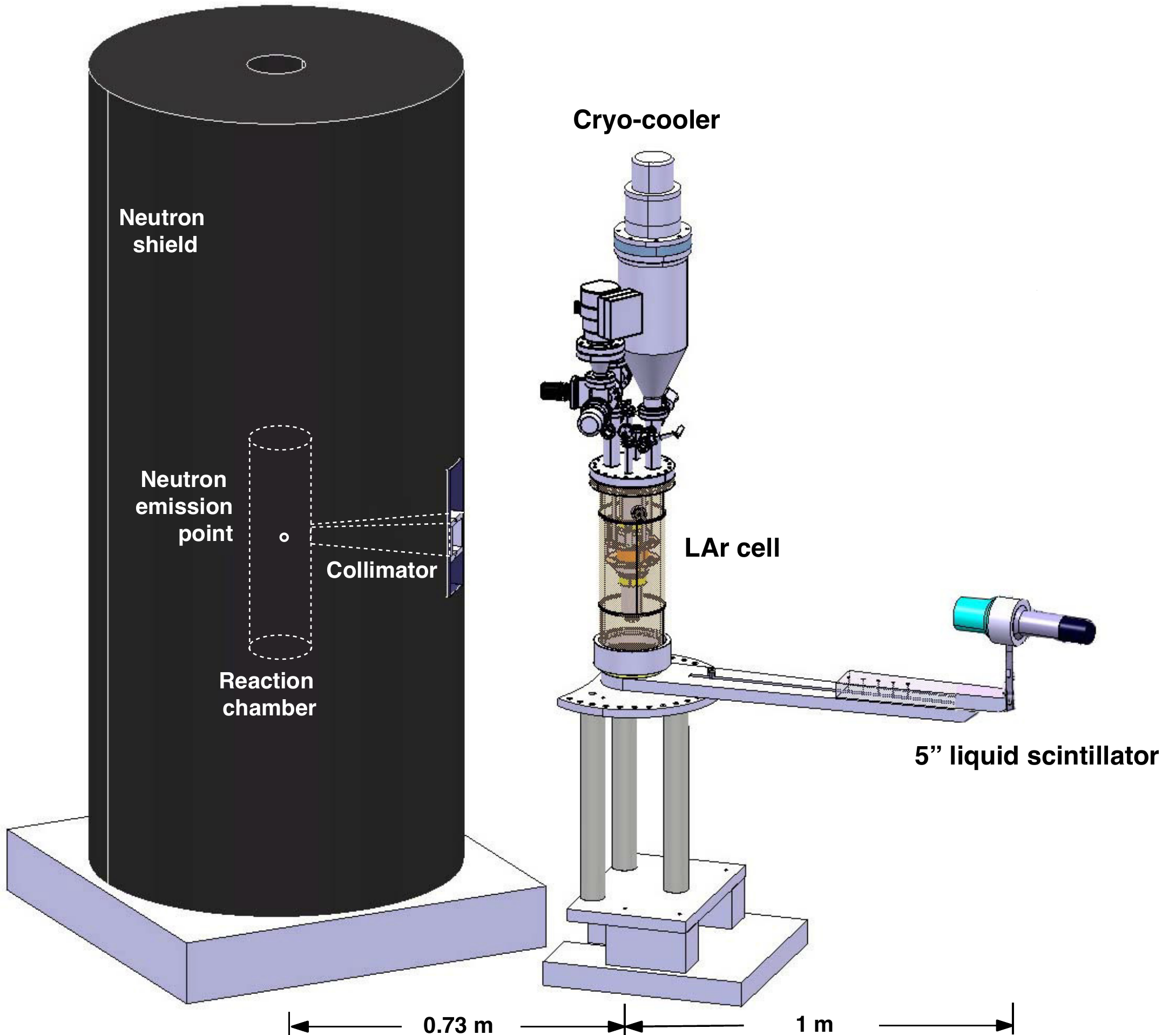}
\includegraphics[width=0.30\textwidth]{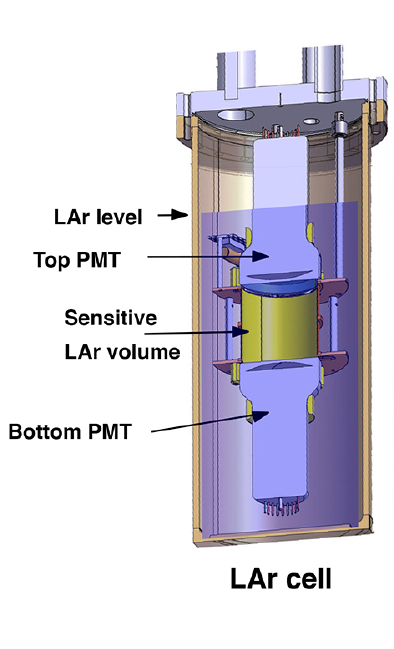}
\caption[]{Left: neutron gun, argon cell and liquid scintillation counter used  to measure $L_{eff}$  in LAr. Right: sketch of the LAr cell with its vacuum chamber (see text).}
\label{Setup}
\end{center}
\end{figure}

We are using  monoenergetic neutrons of $T_n$ = 2.45 MeV from an electrostatic fusion source based on the reaction 
$dd\to ^3$He $n$. The energy deposits are plotted in figure \ref{Principle} (right) as a function of scattering angle. The commercially available neutron generator is  of the deuterium-deuterium plasma fusion type  from NSD-fusion \cite{NSDsite}. The source is a cylinder at ground voltage containing low pressure  deuterium gas ($\mathrm{10^{-2}}$ mbar). An internal perforated cylindrical electrode at high voltage (typically 80 kV) induces a discharge. The ionized deuterons are accelerated towards the inner electrode  and accumulate in the central regions with energies of about 15 keV. This is sufficient to overcome the Coulomb barrier and induce fusion.  The plasma of $\sim$25 mm  length emits neutrons isotropically  into 4$\pi$ with a typical rate of 10$^{6}$ n/s. The certified operation time is 25 000 hours, after which the cell must be exchanged.
The fusion generator is surrounded by a 90 cm diameter shield of borated polyester and the experiment is confined within a radiation controlled fence in our laboratory at CERN  (figure \ref{Setup}, left). Residual radiation (mainly  from scattered neutrons and X-rays)  is well below the authorized limit of 2.5\,$\mu$Sv/h. 

The neutrons are collimated through a polyethylene orifice within roughly 0.2\% $\times$ 4$\pi$ sr to match  the sensitive volume of the LAr cell. The  collimator is defined by a lead box (wall thickness 2 mm) inserted into the polyester to protect against X-rays.  The  LAr cell (figure \ref{Setup}, right) is located at a distance of 73 cm from the neutron emission point. The sensitive volume of the cell consists of a 1.5 mm thin aluminium cylinder (75 mm in diameter and 47 mm high,  hence a LAr volume of 0.2$\ell$). The cell itself is contained in a larger cylindrical vessel (141 mm in diameter), also filled with LAr. The inner wall of the cell is covered by a Tetratex foil coated with a wavelength shifter  made of tetraphenylbutadien (TPB) to shift the VUV scintillation light to a longer wavelength,  following our developments described in  \cite{ArDMPaper}. The surface density of the TPB covering the reflector is 1.0 $\pm$ 0.1 $\mathrm{mg/cm^{2}}$. The light readout is performed with two Hamamatsu R6091-01 photomultipliers (PMT) on each side of the cell. The 3'' bialkali PMTs have 12 dynodes and a Pt underlay to reduce the resistivity of the photocathode when operating at  LAr temperature ($\sim$86 K). The surfaces of the PMTs are coated with a TPB/paraloid  mixture (quantum efficiency estimated to be around 15\%). To perform measurement at zero electric field the aluminium cylinder is polarized at the same voltage as the photocathode of the PMTs. 

The detector is filled with condensed pure argon gas 6.0 containing less than 0.5 ppm of H$_{2}$O, 0.1 ppm of O$_{2}$, 0.1 ppm of H$_{2}$, 0.1 ppm of CO$_{2}$, and 0.1 ppm of CO. Platinum resistors are used to monitor the LAr level inside the cell. The vacuum and gas pressure are read out by pressure sensors. A blue LED ($\lambda$ = 390 nm) connected to an optical fiber is inserted into the vessel for gain calibration. A slow control program written in Labview is used to monitor and record the temperature and pressure during operation.  

For continuous operation a gas system is connected to the cell to provide  condensation and  recirculation of the argon. The gas is condensed on the top of the chamber by the cold head of a Sumitomo CH210 cryocooler with a nominal cooling power of about 80 W, driven by a Sumitomo F-70H helium compressor \cite{Sumitomo}. An outer vacuum chamber at a pressure of  10$^{-5}$ mbar isolates the cryogenic refrigerator from ambient room temperature.  
The purification of the argon gas  is achieved with two purification cartridges OXISORB-W mounted in parallel, which reduce the O$_2$ and H$_2$O levels to  < 5 and < 30 ppb respectively  \cite{OXISORB}. 
The temperature is kept between the solidification and the boiling points of argon (85 K and 87 K respectively at 1 atm) by four resistive heaters located on the first stage cooling station of the cryocooler. Platinum sensors are connected along the coil pipe to monitor the cooling power. A PID controller driven by a LabView program reads out and regulates the temperature of the cold head at the desired value with a precision of about 10 mK. The current provided to the heaters is ensured by a TTi TSX 35 10 programmable PSU based on the GPID interface \cite{TTi}. The typical power produced by the heaters to maintain the temperature for liquefaction is about 50 W. 

The neutrons scattered off the LAr cell are detected by a 5" organic liquid scintillator counter (LSC) manufactured by SCIONIX  \cite{SCIONIX}. The scattering angle $\Theta$ is varied by rotating the LSC around the LAr cell on a  1 m long arm.  The LSC is shielded against direct neutrons from the source by a  10 cm thick polyethylene absorber. Events are recorded by a coincidence measurement of the LAr and LSC signals. To obtain the ratio $L_{eff}$,  equation (\ref{eq:Leff}), the light output produced by the argon nucleus emitted under the corresponding recoil angle is  compared to the reference electron recoil light yield. 

\begin{figure}[hbt]\begin{center} 
\includegraphics[width=0.60\textwidth]{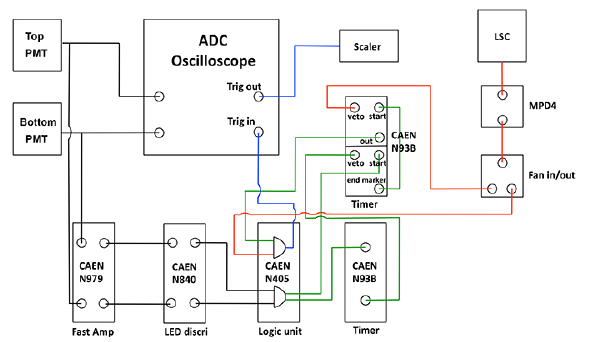}
\caption{Diagram of the data acquisition  system.}
\label{DAQSystem}
\end{center}
\end{figure}

Figure \ref{DAQSystem} shows the schematic of the data acquisition system. The analog signals from the two LAr PMTs are split (not shown in the figure) and fed into  the low and high gain inputs of a 10-bit digitizer (LeCroy Oscilloscope WavePro 735Zi DSO). The low and the high gain signals are matched later offline. This method increases the dynamic range so that signals from single photoelectrons and from $\alpha$-emitters are  collected with high sensitivity without ADC saturation \cite{Regenfusetal}. A cross check is performed by comparing the  pulse height obtained by this method with the pulse height  from several $\gamma$-sources. The accuracy of the signal matching between the high and low gain is $\sim$2$\%$. The data are collected with 5 000 points at 1GS/s with 500 ns pre-sample for pedestal subtraction. Data are stored on a hard disk and then analyzed offline. The signals from each PMT are fed into a CAEN N979  $\times$10 amplifier before being fed into  leading edge CAEN N840 discriminators where the thresholds have been set to 0.5 photoelectrons. The logic signals are then connected to a CAEN N405 unit for the coincidence between the two PMTs.  

\begin{figure}[htb]\begin{center} 
\includegraphics[width=0.65\textwidth]{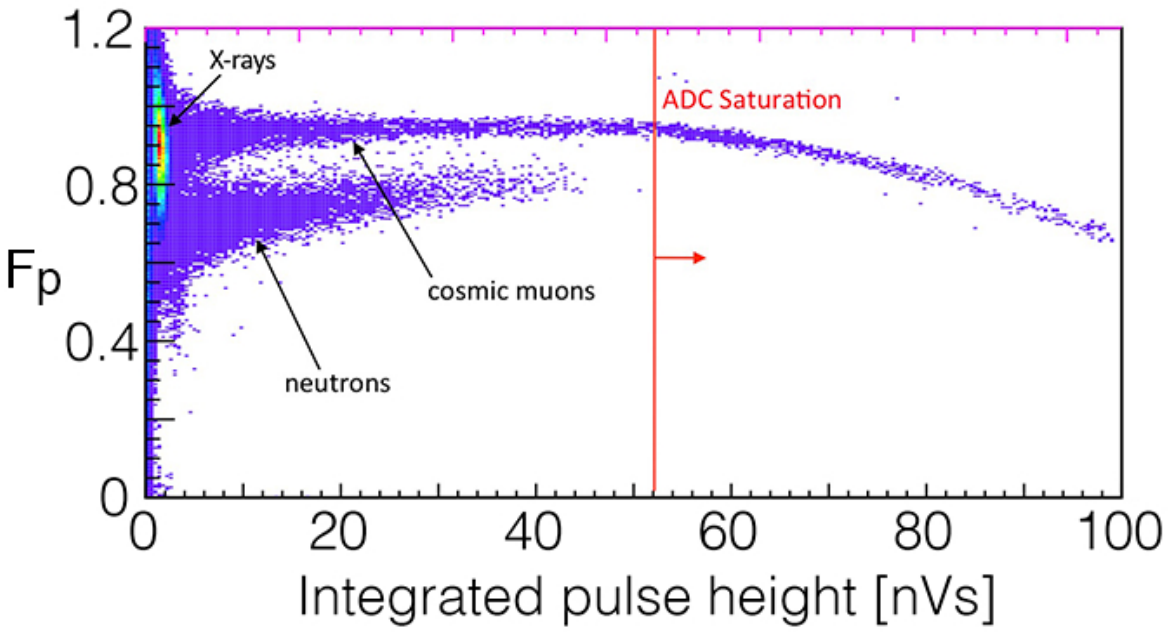}
\caption{\label{FpLSC} Prompt fraction $F_p$ of scintillation light in the LSC versus integrated pulse height. The upper band is produced by electronic recoils from bremsstrahlung (generated by the neutron generator) and cosmic muons, the lower band by proton recoils from neutron scattering.}
\end{center}
\end{figure}

Neutron signals from the LSC are processed by an analog pulse shape discriminator module Mesytec MPD4 \cite{MPD4}. This module determines the fraction of fast ($<$20 ns) component of the scintillation light. For electrons the prompt fraction $F_p$ (prompt over  total integrated pulse height) is about 0.95, while for proton recoils $F_p < $ 0.8 (figure \ref{FpLSC}). 

The energy calibration of the LSC is performed with radioactive $\gamma$-sources. The energy resolution is determined from the signals induced by backscattered photons from   $^{137}$Cs or   $^{22}$Na sources, which are detected by a small BaF crystal \cite{Dietze}. 
The time-of-flight between the LAr cell and the LSC is measured to select elastic events. The LSC signal is fed into a linear fan in/out module which feeds the timer and logic units. The time-of-flight between the LAr cell and the LSC is calculated from the time difference of the TAC output of the MPD4 module and the signal from the LAr cell. The time-of-flight is  calibrated by the coincidence of the  511 keV  back-to-back $\mathrm{\gamma}$-rays from a $^{22}$Na source located at 
mid-distance between the LAr cell and the LSC. Events are accepted within a time delay of 200 ns between the  LAr and  LSC signals.  This trigger setting was used during part of the data taking (TR$_2$ data sample, see table \ref{NGHistoryDataTaking} below). For the TR$_1$  data sample taken earlier a programmable trigger logic in the oscilloscope was used for the coincidence between the  LAr and LSC signals.

\section{Calibrations with $\gamma$-sources}
\subsection{Pulse height analysis}
The light yield from the LAr cell is  the ratio of  collected to produced light intensities, which strongly depends on factors such as geometry, photomultiplier efficiency, wavelength shifter, etc. The energy calibration in LAr is performed   by using the 60 keV photo-absorption peak from an external $\mathrm{^{241}Am}$ source. The source is placed at a distance of 10 cm from the vessel  to reduce pile-up events. The light yield has to be corrected for the finite integration time. As mentioned earlier, nuclear recoils contribute mostly to the fast component ($\tau_1$), while the contribution from electrons  to the slow component ($\tau_2$) is larger. Therefore the light yield  has to be corrected first for losses due to impurities  in LAr which are reducing the lifetime of the slow component \cite{Amslerquench}. 

\begin{figure}[htb]
\begin{center}
\includegraphics[width=0.54\textwidth]{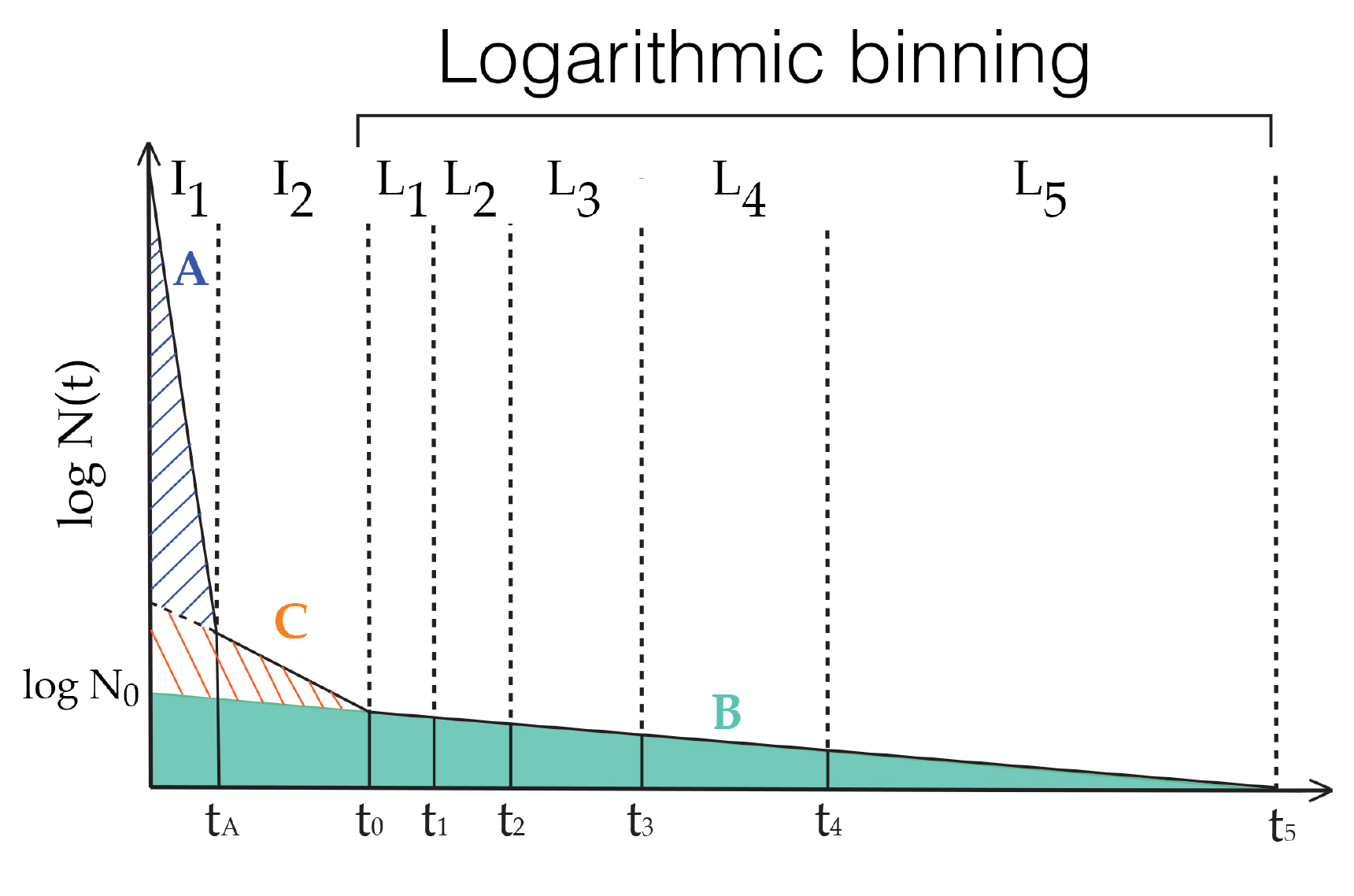}
\includegraphics[width=0.45\textwidth]{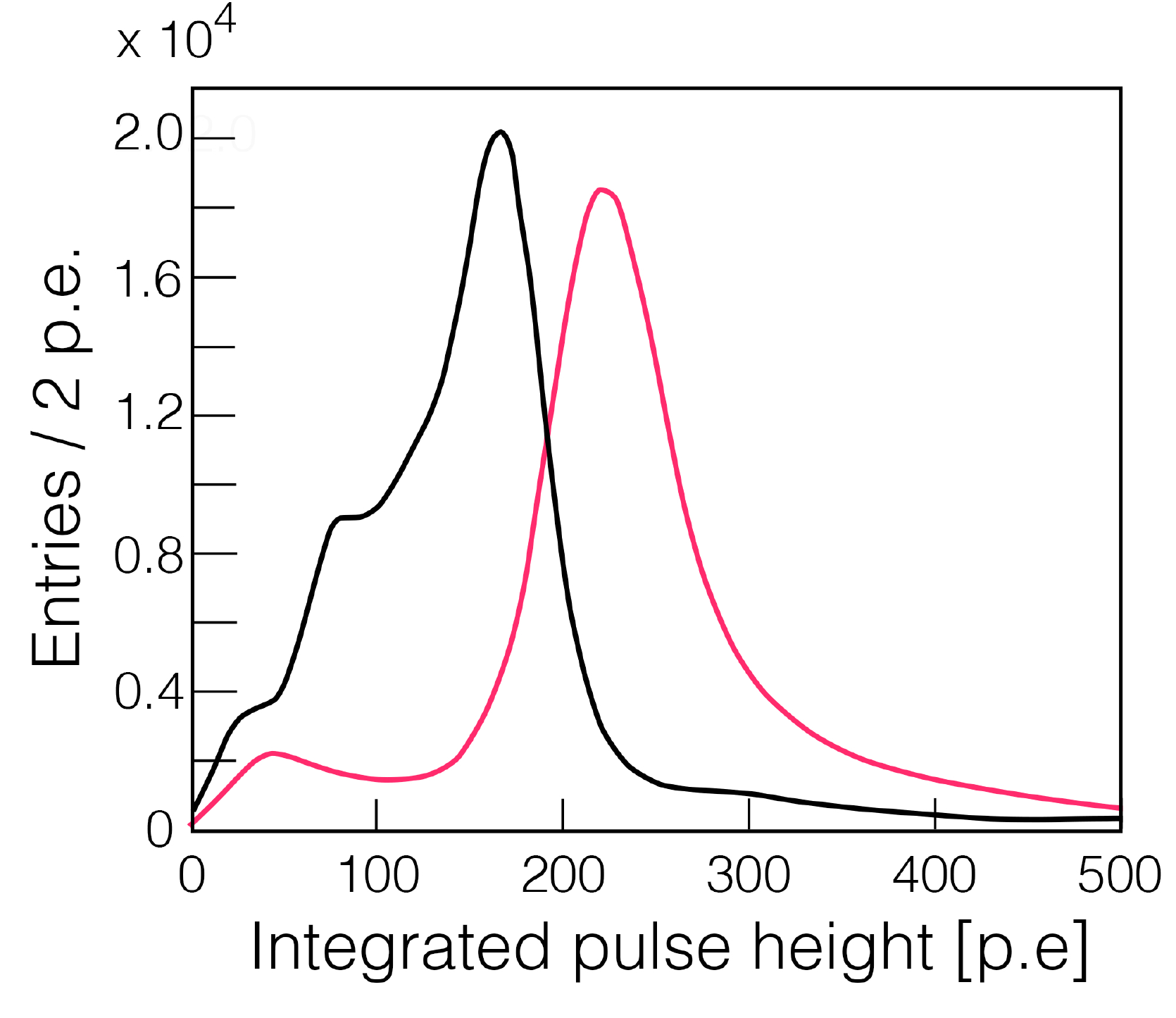}
\caption[]{Left: logarithmic binning method showing the contributions of the  populations A, B and C. Right: pulse height distribution from the 60 keV $\gamma$-source (in photoelectrons, p.e.) for the raw data (black curve) and after correction with the log binning method  (red curve).
\label{LogBinScheme}}
\end{center}
\end{figure}

We have developed a new technique to correct for the reduced light yield caused by argon impurities. This event-by-event method consists in dividing the pulse shape in three time regions $I_1$, $I_2$, and $L$, as sketched in 
figure~\ref{LogBinScheme} (left). Time $t = 0$ corresponds to --20 ns below the maximum pulse height. The first interval  ($I_1$)  corresponds to 4$\tau_{1}$. The integral over $I_1$ contains the contributions $A$ from the fast component and $B$ from the slow component. A third contribution $C$ is present in the interval $I_2$, between 40 ns and 100 ns. This component has been observed earlier in argon and is reported to stem from the wavelength shifter  \cite{Segreto}.  The rest of the pulse shape  is divided logarithmically into $n = 5$ time bins $L_i$ from $t_{0}$ to $t_{5}$ defined as
\begin{equation}
t_i = t_0 + \tau_{2}\cdot\ln\left(\frac{n+1}{n+1-i}\right)
\end{equation}  
with $i = 1...5$ and where $\tau_{2}$ is the measured decay time of the slow component. Thus the expected number of events in each bin $L_i$ is constant. Impurities affect the late regions of the light pulse, hence affect only the slope of the slow component $B$  but not its amplitude $N_0$ at $t = 0$. The measured number of entries in $L_i$ can be obtained by building the arithmetic mean of the five bins. The contribution $B_{corr}$ from the slow component, corrected for impurities, is then obtained by integrating $N(t)$ with the time constant $\tau_2$ = 1.6 $\mu$s which corresponds to pure LAr. Once $B_{corr}$ is obtained, $C$ is found from the $I_2$ integral and, finally, $A$ from the $I_1$ integral after subtracting the contribution from $B_{corr}$ and $C$. Details can be found in \cite{Creus}. Let us define the fraction of prompt light determined with the logarithmic binning by the ratio
\begin{equation}
r_{LB} = \frac{A}{A+B_{corr}+C}.
\label{eq:rLB}
\end{equation}

Figure \ref{LogBinScheme} (right) shows the raw  energy distribution from  the  $^{241}$Am source  for runs taken under various purity levels.  Mixing data with different purities leads to a spread of the deposited energy distribution (black curve). However, applying the logarithmic  binning method to correct the spectrum leads to a  decrease of the width of the distribution and shifts the distribution towards higher  photoelectron numbers.  

\begin{figure}[htb]
\begin{center} 
\includegraphics[width=0.5\textwidth]{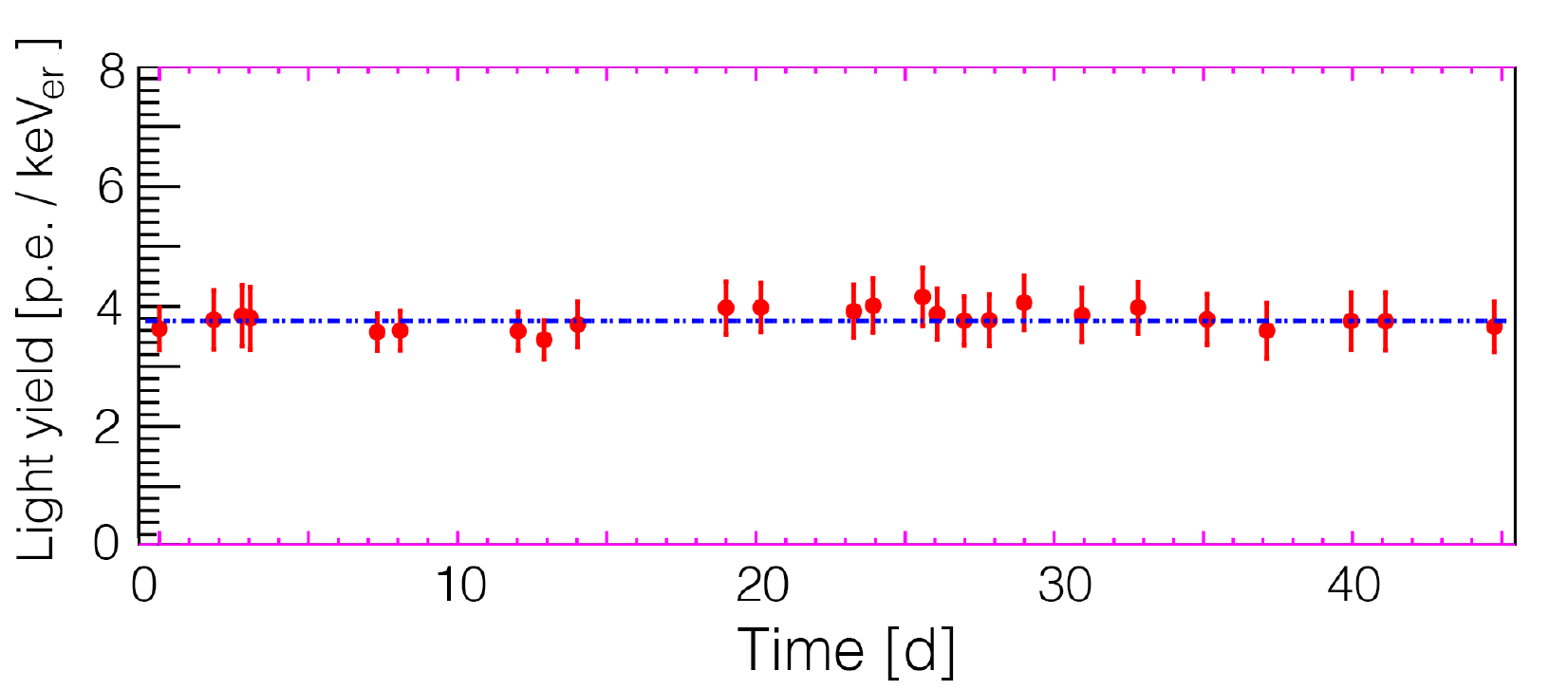}
\caption[]{Light yield as a function of time for the TR$_1$ data sample.}
\label{60keV}
\end{center}
\end{figure}

\begin{figure}[htb]\begin{center} 
\includegraphics[width=0.38\textwidth]{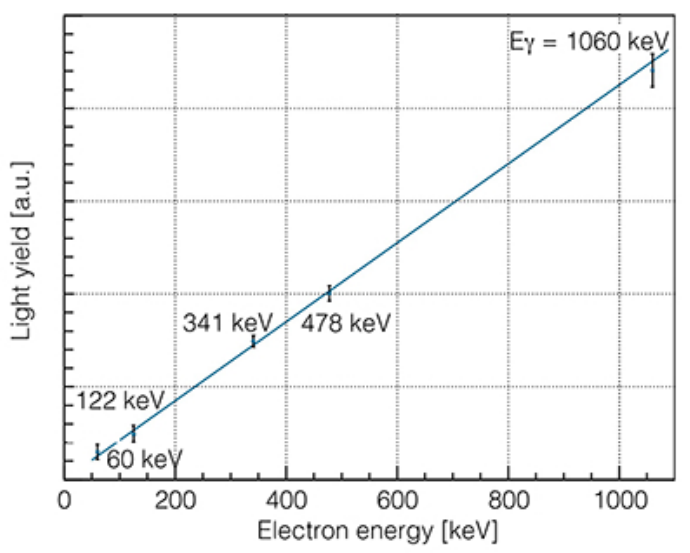}
\caption{\label{VariousSources} Light yield from LAr measured with various $\gamma$-sources (in arbitrary units). In the present work the calibration is obtained from the light yield of the 60 keV  $^{241}$Am photopeak.}
\end{center}
\end{figure}

Light yield calibrations were performed periodically during  data taking. The corrected average light yields are $3.75 \pm0.08$ p.e./keV$_{er}$ and $3.39\pm0.07$ p.e./keV$_{er}$ for the TR$_1$ and TR$_2$ data respectively  \cite{Creus}. The slightly different light yields are due to the different thicknesses of wavelength shifter covering the surfaces of the PMTs. The long term stability of the light yield is shown in figure \ref{60keV}  for the TR$_1$ data. Further external radioactive sources were employed to cover a wide energy range: 122 keV photoelectron from $^{57}$Co, 341 keV, 478 keV and 1060 keV Compton edges from $^{22}$Na,  $^{137}$Cs and $^{22}$Na respectively. Figure \ref{VariousSources} illustrates the perfect linearity of the system. 

The response to very low energy electrons was also measured with a $^{83}$Rb source (half-life of 86 days) of about 180 kBq which was  introduced into the system. A $^{83}$Rb trap was directly connected to the gas system. The daughter metastable $^{83\mathrm{m}}$Kr nuclide (half-life of 1.8 h)  illuminates the central part of the LAr cell and  diffuses photons uniformly. The measured light yield at 32.1 keV is obtained as $3.68\pm0.15$ p.e./keV$_{er}$ in good agreement with results from the other sources.

\subsection{Trigger efficiency}
\label{sec:trigeff}
\begin{figure}[htb]
\includegraphics[width=0.55\textwidth]{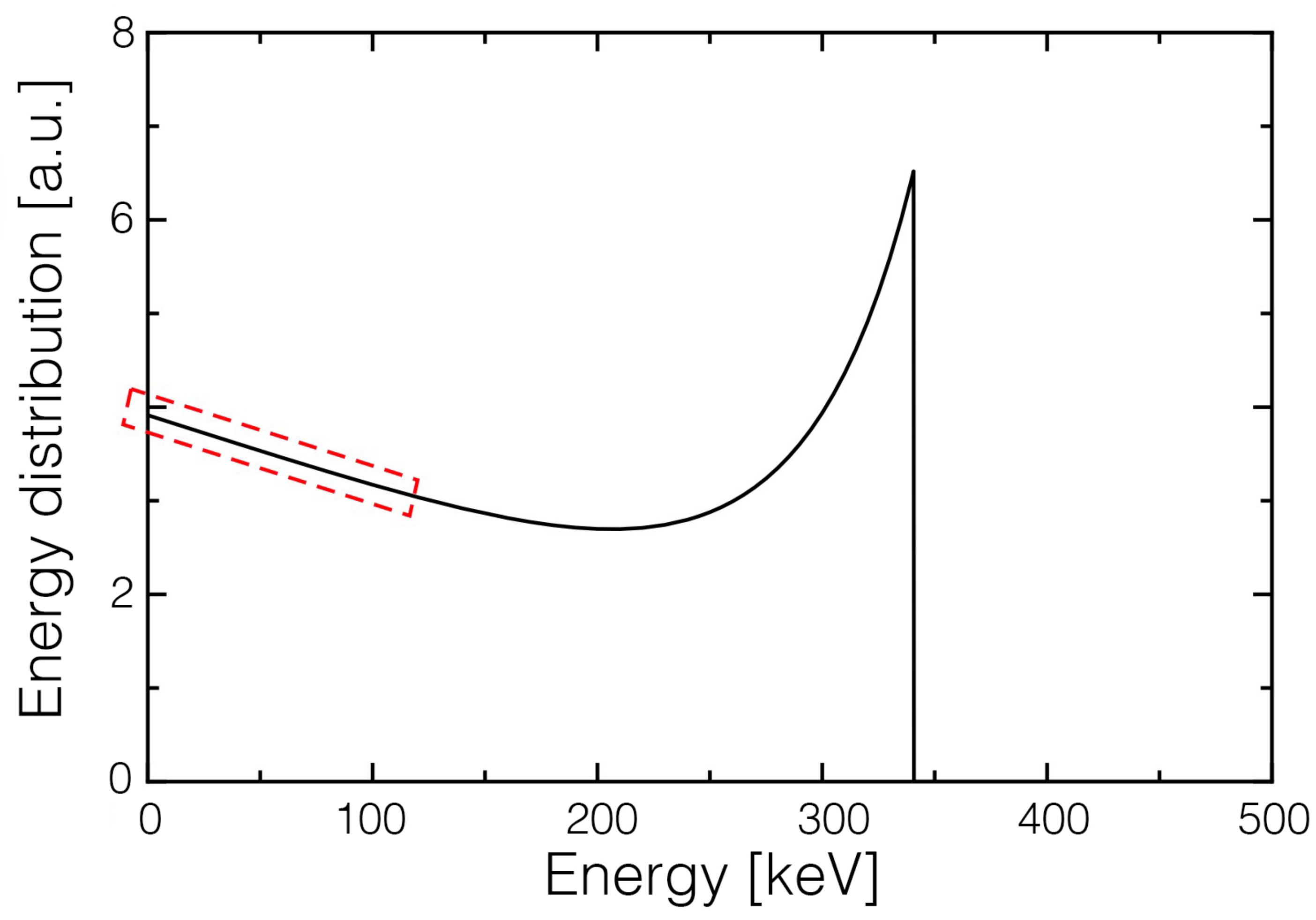}
\includegraphics[width=0.44\textwidth]{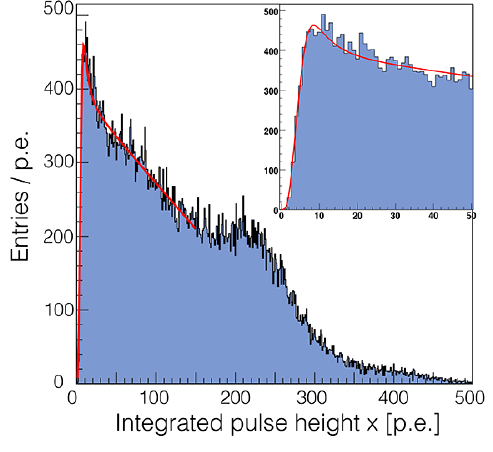}
\caption[]{Left: theoretical electron recoil spectrum from the 511 keV annihilation $\gamma$ of a $\mathrm{^{22}Na}$ positron source  (computed with the Klein--Nishina formula). The red dashed box shows the linear part of the spectrum at low energy. Right: measured integrated pulse height $x$ of the prompt light in photoelectrons (p.e.) fitted  with the function (\ref{ComptFitFormula}) (red curve). The inset shows the low energy region.
\label{PlotComptonfit}}
\end{figure}

Understanding the low energy roll-off of the trigger efficiency is essential to determine $L_{eff}$, especially  for low energy recoils. We have studied the trigger efficiency with the 511 keV annihilation $\gamma$-rays from a $\mathrm{^{22}Na}$ positron source. The energy distribution of the Compton scattered electron  is linear at low energy, as illustrated by the red dashed box in 
figure \ref{PlotComptonfit} (left). As a comparison figure \ref{PlotComptonfit} (right) shows the measured electron recoil spectrum in LAr. 

\noindent
The low energy part  is fitted with the following model function: 
\begin{equation}
\mathcal{F}(x)= S\cdot [(D+E\cdot x)+L(a,\mu,\sigma)].
\label{ComptFitFormula}
\end{equation}
The function
\begin{equation}
S = (1-\mathrm{e}^{-\frac{x}{b}})^{c}
\label{eq:S}
\end{equation}
describes the  trigger efficiency, where $x$ is the pulse height in photoelectrons. The term $(D+E\cdot x)$ describes the linear shape of the spectrum at low recoil energies. The parameters $D$ and $E$ are determined by the fit. The Landau distribution $L(a,\mu,\sigma)$ takes the \v{C}erenkov light from the $\gamma$-ray interaction in the PMT glass into account, where $a$ is a normalization parameter, $\mu$ the most probable value and $\sigma$  a scale parameter (adapted from the CERNLIB routine G110 {\it denlan}).  Most of the \v{C}erenkov events contribute to the prompt light. 

\begin{figure}[htb]\begin{center} 
\includegraphics[width=0.40\textwidth]{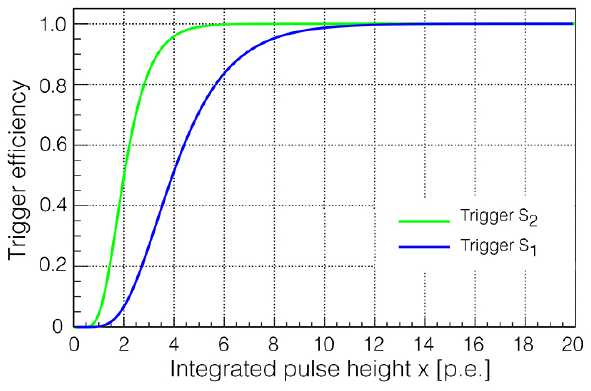}
\caption{\label{TriggerCurve}Trigger efficiency $S$ for the  $\mathrm{^{22}Na}$ source measurement as a function of  integrated pulse height $x$  for the two  trigger settings.}
\end{center}\end{figure}

The fit is performed in two steps. First, the linear part of the spectrum is fitted. Second, the parameters of the linear function and the Landau function are fixed and the parameters $b$ and $c$ are determined.  The measured trigger efficiency $S$ is displayed in figure \ref{TriggerCurve} by the green curve for the S$_2$ data. The efficiency is $\sim$$100\%$ at 8 photoelectrons (p.e.) , $\sim$$96\%$ at 4 p.e., $\sim$$50\%$ at 2 p.e. and $\sim$$5\%$ at 1 p.e. The roll-off of the trigger efficiency for the TR$_1$ data that used the programmable trigger logic in the oscilloscope is shown by the blue curve. The efficiency is lower, reaching $\sim$$95\%$ at 8 p.e., $\sim$$55\%$ at 4 p.e. and $\sim$$6\%$ at 2 p.e.

\section{Neutron data}
The neutron generator was operated most of the time at 80 kV to keep the accidental background from bremsstrahlung generated by the neutron generator at an acceptable level.   The corresponding neutron flux is 6 $\times$ 10$^{5}$ neutrons/s into 4$\pi$. The scattered rate from the sensitive LAr volume, recorded by the LSC, is 1 neutron/min. The background rate is estimated to be  5/min, caused mainly by diffusely scattered neutrons and X-rays from bremsstrahlung ($\sim$3--4/min), and to a lesser extent by cosmic muons saturating the MPD4 analog pulse shape discriminator.

\begin{table}[htb]
\begin{center}
\begin{tabular}{cccccc}
\hline
$\mathrm{\Theta}$&$\mathrm{T_{nr}}$[keV]&Trigger&Running&Trigger&Running\\
$[^\mathrm{o}]$ & [keV]& &time [10$^3$s]& &time [10$^3$s]\\
 \hline
25&\ \ 11.5&TR$_1$ & 269 &TR$_2$ & 382\\
30&\ \ 16.4&TR$_1$ &284 &TR$_2$ &234 \\
40&\ \ 28.5&TR$_1$ &379 &TR$_2$ &219 \\
50&\ \ 43.4&TR$_1$ &291&--&-- \\
60&\ \ 60.5&TR$_1$ &288 &--&--\\
90&119.5&TR$_1$ &196 &--&--\\
 \hline
 \end{tabular}
\caption{Angular settings and corresponding argon recoil energies. The trigger type (see text) and running time are also given.}
\label{NGHistoryDataTaking}
\end{center}
\end{table}

Neutron scattering data are measured at six angles corresponding to argon recoil energies  between 11 and 120 keV (table \ref{NGHistoryDataTaking}).  The data are taken with the two different triggers TR$_1$ and TR$_2$.  The raw data are first corrected for  impurities and for the trigger efficiency roll-off described in section \ref{sec:trigeff}. 

\begin{figure}[htb]
\begin{center}
{\includegraphics[width=0.42\textwidth]{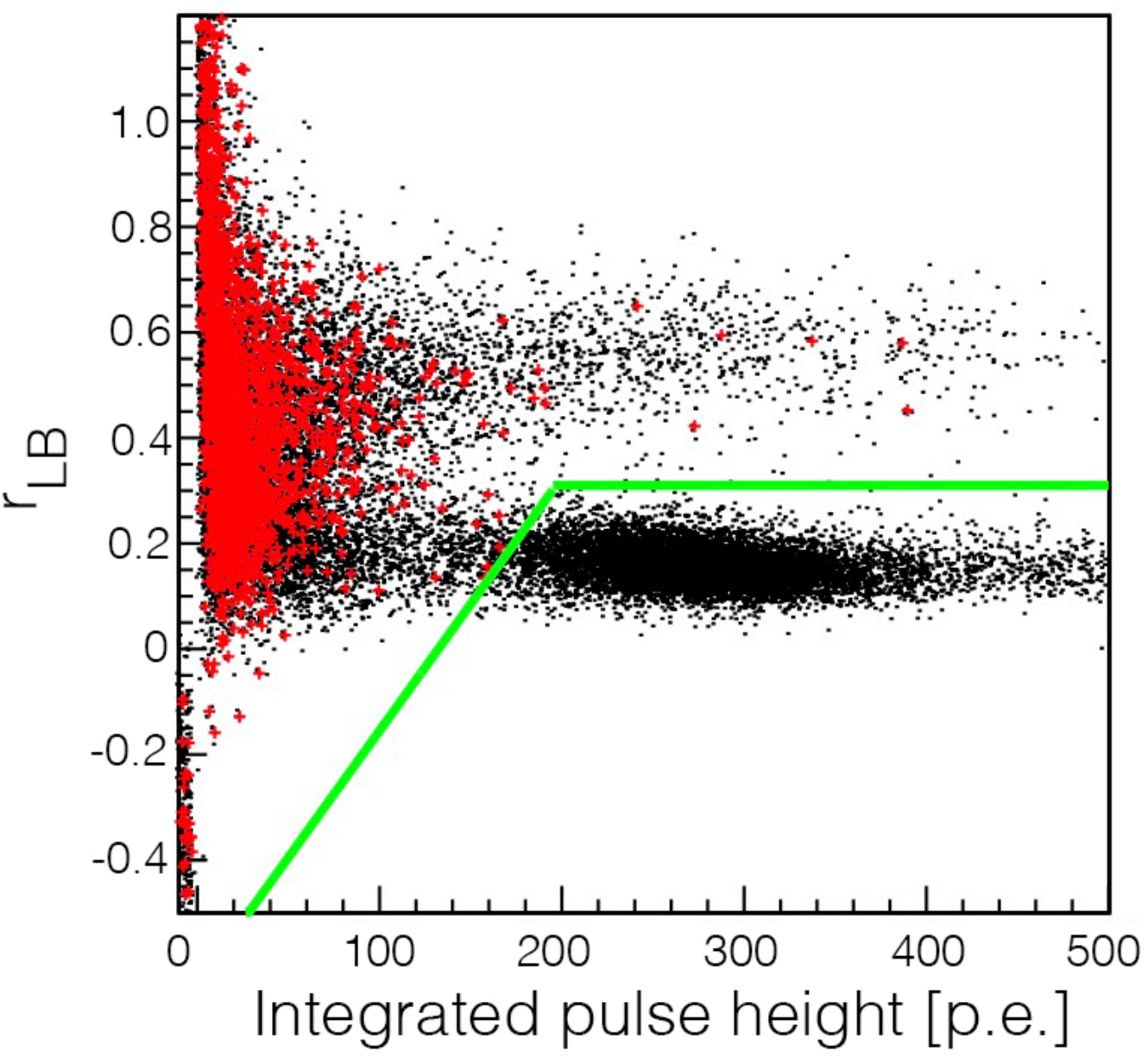}}
\caption{\label{LBSIPH} Scatterplot of the $r_{LB}$ discriminator  versus  integrated pulse height for $\Theta$ = 25$^\mathrm{o}$. The green line shows the cut applied to select elastic neutron scattering (red region) and reduce background.}
\end{center}
\end{figure}

The discriminator $r_{LB}$  (\ref{eq:rLB}), which is based on the logarithmic binning method (and thus differs from the prompt fraction $F_p$), suppresses background from accidental coincidences with X-rays, $\gamma$-rays  and cosmic muons. As an example, figure \ref{LBSIPH} shows a scatterplot of $r_{LB}$ in LAr as a function of integrated pulse height for the $\Theta$ = 25$^\mathrm{o}$ data sample. Events above the green line are mostly neutron induced, those below the line stem mostly from background.

\begin{figure}[htb]
{\includegraphics[width=0.99\textwidth]{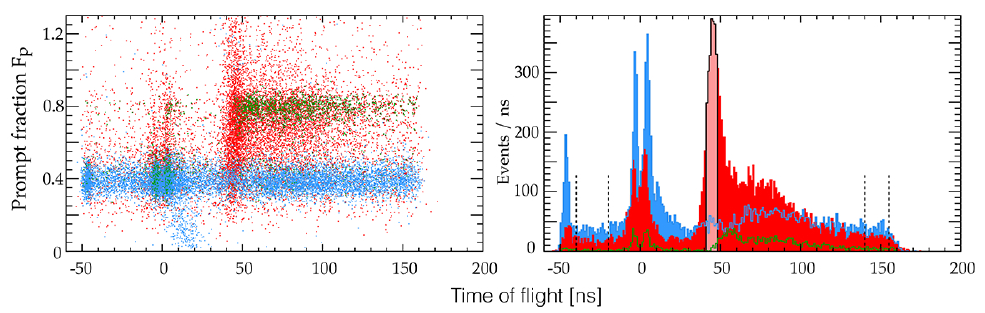}}
\caption{\label{TOFPlotS$_1$} Prompt fraction versus time-of-flight  (left)  and corresponding projection on the time-of-flight axis (right) for 
$\Theta$ = 25$^\mathrm{o}$. The events in red correspond to nuclear recoils selected by the $r_{LB}$ cut, while the anticut selects  the electron-recoils (events in blue). The events in green correspond to nuclear recoils at energies above 120 photoelectrons. The  elastic neutron scattering events selected to calculate $L_{eff}$  lie in the pink window. The vertical dashed lines delimit the intervals used to evaluate the contribution from accidental events.}
\end{figure}

Figure \ref{TOFPlotS$_1$} (left) shows a scatterplot of the prompt light fraction $F_{p}$ (defined in section \ref{sec:Fp}) versus  time-of-flight  for $\Theta$ = 25$^\mathrm{o}$.  The event distributions for the data taken at the larger scattering angles are similar. 
The time-of-flight projection is shown in figure  \ref{TOFPlotS$_1$} (right).  Events in red are  nuclear recoils, while events in blue are mostly due to background  (lying below the green line in figure \ref{LBSIPH}). Since the LAr  and the LSC triggers are not correlated the accidental background is uniform and appears as a flat component in the time-of-flight spectrum.

The most prominent red peak  is due to elastically scattered neutrons at the expected time-of-flight of about 43 ns. These events are also associated with large values of $F_p$ and therefore contribute mostly to the prompt light. An accumulation of events is observed around 80 ns, corresponding to inelastically scattered neutrons in LAr. The broad tail of this distribution is mainly due to neutron scattering outside the active volume and inelastically scattered neutrons.
The two peaks around zero time-of-flight are due to cosmic muons flying between the LAr cell and the LSC, the peak at --3.1 ns corresponding to muon from the LSC to the LAr cell and the one at +3.1 ns to those flying in the opposite direction. 
In the electronic (blue) recoil band an accumulation of events is observed above the elastic peak. They are correlated in time with neutrons  scattered inelastically in the target but outside the  sensitive volume, while the ensuing emitted photons are converted in the active volume. 
The events shown in dark green are nuclear recoil events with energy deposits  above 120 photoelectrons, mostly due to multiple neutron scattering. The accumulation of events at --50 ns is due to an electronics artifact at the edge of the trigger time window.

\begin{figure}[htb]
\begin{center} 
{\includegraphics[width=7cm,angle=0]{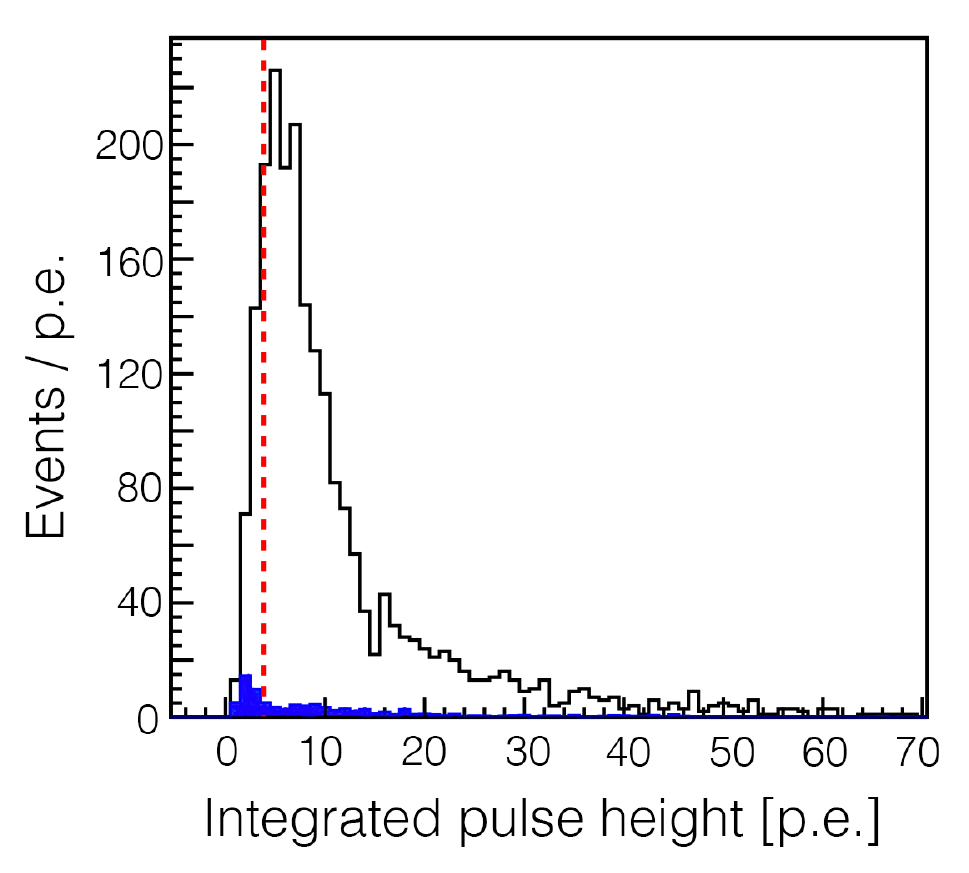}}
\caption{\label{EnergyRecoil1} Nuclear recoil energy distributions  at $\Theta$ = $25\mathrm{^{o}}$ (in photoelectrons). The plot shows the contribution to the fast light. The accidental spectrum is shown in blue. The vertical red dashed line indicates the pulse height corresponding to 90\% trigger efficiency.}
\end{center}
\end{figure}

Two cuts are applied to select elastic neutron scattering. A time-of-flight cut window is applied between --2 ns and +6 ns of the most probable value for the elastic peak  (pink area in figure \ref{TOFPlotS$_1$} corresponding to 41 -- 48 ns at $\Theta$ = 25$^\mathrm{o}$).  The cut by  the   $r_{LB}$ discriminator  (green line in figure \ref{LBSIPH}) removes efficiently accidental coincidences with X-rays, $\gamma$-rays, inelastic scatters and cosmic muons. Figure \ref{EnergyRecoil1} shows the contribution to the fast light  for the selected data.   The recoil energy distribution shown in this figure will be fitted to the Monte Carlo data, as described in section \ref{sec:simulation}. 

To estimate the contribution from residual accidentals we choose events lying far from the elastic peak in the two intervals delimited by the vertical dashed lines in figure \ref{TOFPlotS$_1$}. The background spectra below and above the elastic peak are found to be compatible. The contribution from accidentals is negligible, as shown by  the blue histogram in figure \ref{EnergyRecoil1}. However, neutrons that scatter off various materials outside the LAr cell before or after interacting in the LAr sensitive volume also contribute background. This external  background cannot be removed from the data  and is therefore taken into account by Monte-Carlo simulation. 

\section{Monte-Carlo simulation}
\label{sec:simulation}
\begin{figure}[!h]\begin{center} 
\includegraphics[width=0.40\textwidth]{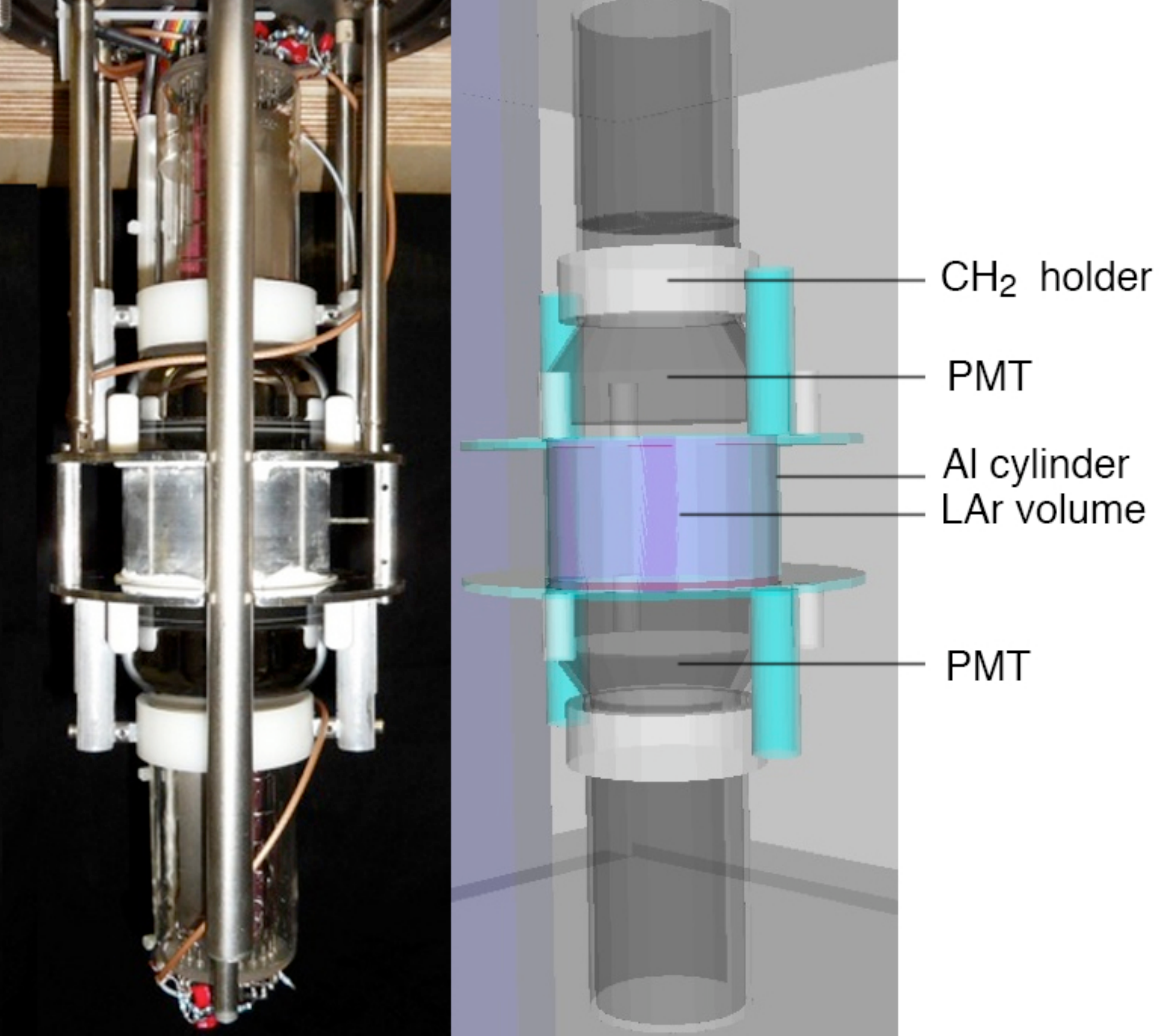}
\caption{\label{MCGeometry} Photograph of the LAr cell and PMT  (left) and drawing of its simulated counterpart (right).}
\end{center}\end{figure}

The detector response was simulated extensively with GEANT4. The Monte-Carlo simulation takes into account the neutron generator assembly with its shielding and collimator, the cryostat and the various components of the LAr cell such as the vacuum vessel, the TPB reflector foil, the photomultipliers, the support mechanics, and the LSC. A photograph of the LAr cell and a drawing of the simulated one are compared in figure \ref{MCGeometry}. 

\begin{figure}[!h]\begin{center} 
\includegraphics[width=0.80\textwidth]{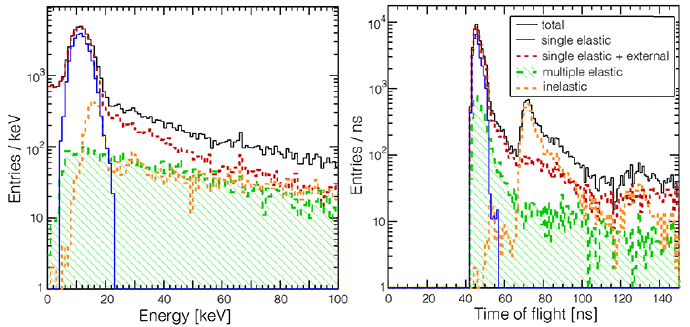}
\caption{\label{25DegSim} Simulated recoil energy distribution at $\Theta$ = 25$^\mathrm{o}$ (left) and corresponding time-of-flight spectrum (right) from the contributions listed in the inset: single elastic scattering (with and without external scattering), multiple scattering and inelastic scattering in the LAr sensitive volume.}
\end{center}
\end{figure}

In the simulation monoenergetic neutrons of 2.45 MeV are emitted isotropically into 4$\pi$. A sample of 100 million interactions in the active LAr volume was generated at each measured angle. 
Figure  \ref{25DegSim} (left) shows for $\Theta$ =  25$^\mathrm{o}$ the various simulated contributions to the  nuclear recoil energy spectrum (left) and to the time-of-flight distribution (right) prior to smearing by the experimental resolution. A clear elastic peak is observed at the  expected recoil energy, as well as the exponentially decreasing material background tail  (black histogram). Neutrons that scatter elastically only once in the active  LAr volume are shown by the blue histogram. The dashed red histogram corresponds to single scattering in the active LAr volume, preceded or followed by scattering elsewhere in the apparatus (external background). The multiple scattering events in LAr are shown in green.  Inelastically scattered neutrons in LAr  are represented by the  dashed orange histogram. The  energy deposits are shifted to higher values by the conversion of the $\gamma$-rays emitted  during the de-excitation of the argon atoms. 

Figure  \ref{25DegSim} (right) shows the corresponding time-of-flight distributions at $\Theta$ = 25$^\mathrm{o}$. The position in time of the elastic peak is consistent with data.  A comparison with the data of figure \ref{TOFPlotS$_1$} shows that nearly all  inelastic scatters are removed by the time-of-flight cut  of 41-- 48 ns. Their time-of-flight distribution (enhancement around 70 ns) is also in good agreement with data. 

\begin{figure}[htb]
\begin{center}
\includegraphics[width=0.4\textwidth]{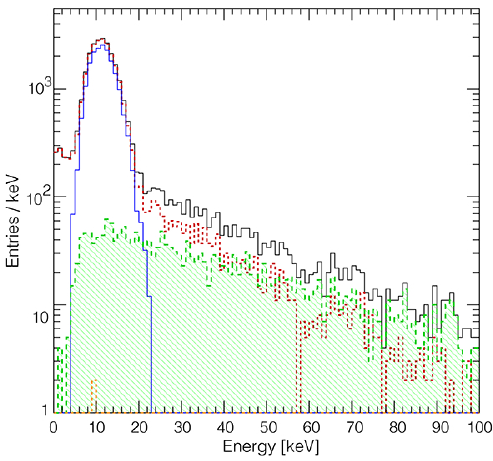}
\caption[]{Simulated recoil energy distribution at 
$\Theta$ = 25$^\mathrm{o}$ after the time-of-flight cut. The color codes are the same as in figure \ref{25DegSim}.
\label{25DegSpectSimTOFcut}}
\end{center}
\end{figure}

The position of the single elastic peak and its resolution depend on the sizes of the neutron source and on the solid angle acceptances of the LSC and  LAr cell. Uncertainties on recoil energies are determined by fitting a Gaussian to the single elastic peak. Before applying the time-of-flight cut to the simulated data the finite time resolution is taken into account by convoluting a Gaussian with the experimental rms resolution of 2.88 ns. The recoil energy distribution at $\Theta$ =  25$^\mathrm{o}$ is shown in figure \ref{25DegSpectSimTOFcut} after the time-of-flight cut. The black histogram will be used to fit to the data of figure \ref{EnergyRecoil1}. 

The various contributions to the nuclear recoil spectrum after the time-of-flight cut are listed in table \ref{ContrClassEvts} for the six scattering angles.  Below 40$^\mathrm{o}$ single elastic scattering off argon is the dominant contribution to the recoil energy spectra.  The contribution from multiple scattering in LAr is small ($\sim$$10\%$) thanks to the small dimension of the LAr cell compared to the elastic neutron mean free path. The decreasing contribution of elastic events with increasing scattering angle is due to the angular dependence of the differential elastic cross section for neutron-argon interaction at 2.45 MeV which reaches a minimum around 90$^\circ$ \cite{NNDC}.  

\begin{table}[!h]
\begin{center}
\begin{tabular}{ccccc}
\hline
$\mathrm{\Theta}$&Single &Single elastic &Multiple &Inelastic \\
& elastic & + external & elastic &\\
\ [$^{\mathrm{o}}$]&[$\%$]& [$\%$]& [$\%$]& [$\%$]\\
\hline
25&67.7&23.5&\ \ 8.8&0.02\\
30&65.3&25.1&\ \ 9.6&0.02\\
40&54.8 &31.4&13.7&0.06\\
50&49.7&34.3&16.0&0.04\\
60&40.9&31.8&20.5&0.04\\
90&29.1&45.1&25.6&0.18\\
 \hline
 \end{tabular}
\caption{\label{ContrClassEvts} Rms contributions to the recoil spectrum after time-of-flight cut.}
\end{center}
\end{table}

\section{Data analysis}
\subsection{$\chi^2$-fits}
\label{finalLeffResults}
The method developed in this work to extract $L_{eff}$ follows the one described in \cite{GPlante,PlanteThesis} for liquid xenon. We have performed a $\chi^{2}$-minimization  to determine the relative light efficiency $L_{eff}$ (\ref{eq:Leff}) as a function of recoil energy  $T_{nr}$  at   six scattering angles. 

\begin{figure}[htb]\begin{center} 
{\includegraphics[width=0.9\textwidth]{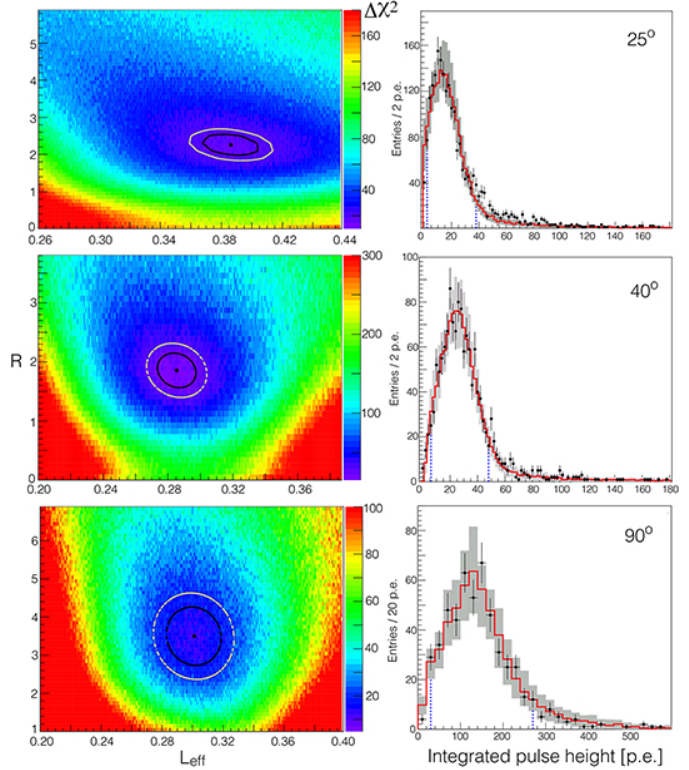}}
\caption{\label{FitParabMCvsData}Left: $\chi^{2}$ maps as a function of $L_{eff}$ and $R$ for three scattering angles. The location of the minimum $\chi^{2}$ is shown by a dot. The curves are the 1$\sigma$ and $2\sigma$ contour lines. Right: corresponding Monte-Carlo fits (red histograms) to the energy recoil spectra  (black data points) with rms errors (grey zones). The dotted blue lines delimit the fitted ranges.}
\end{center}
\end{figure}

To obtain the simulated spectrum in photoelectrons the energies in keV are first multiplied by the free parameter ${L}_{eff}$  under $\chi^{2}$ test and then by the light yield $Ly$ measured with the $^{241}$Am source. 
The number of photoelectrons $N_{p.e.}$  is allowed to fluctuate according to a Poisson distribution and the energy resolution is described by the rms deviation of a Gauss distribution defined as $R\sqrt{N_{p.e.}}$, where $R$ is the second  free  parameter under $\chi^{2}$ test. The gain fluctuation of the PMTs is taken into account by convoluting the nuclear recoil energy spectrum in photoelectrons with a Gaussian distribution with rms deviation  $R_{p.e.}\sqrt{N_{p.e.}}$. The parameter $R_{p.e.}$ is determined from the resolution of the measured PMT single photoelectron distribution and found to be $R_{p.e.} = 0.4$ \cite{Creus}. The trigger efficiency (figure \ref{TriggerCurve}) is then taken into account to obtain the simulated recoil energy spectrum. 
For each scattering angle the $\chi^{2}$ is computed as follows:
\begin{equation}
\chi^{2}(L_{eff},R)=\sum_{i=1}^N \frac{[h_{i}-h_{MC,i}(L_{eff},R)]^2}{\sigma_{i}^{2}+\sigma_{MC,i}^{2}},
\label{eq:chisq}
\end{equation}
where  $L_{eff}\equiv L_{eff}(T_{nr})$ and the resolution $R\equiv R(T_{nr})$ are the free parameters. The events are divided into $N$ bins expressed in photoelectrons. The number of events in the measured and the simulated bins  $i$ are labelled $h_{i}$ and $h_{MC,i}$ respectively. The corresponding statistical uncertainties are denoted by $\sigma_{i}$ and $\sigma_{MC,i}$. The fit range is chosen separately for each angle in such a way as to avoid the bias induced at low energy by the trigger inefficiency (which underestimates $L_{eff}$) and by the high energy tail (which overestimates $L_{eff}$). Fits are performed for different photoelectron ranges and the suitable fit range is selected as the one associated with stable values of $L_{eff}$. For example, at 25$^\mathrm{o}$  the range fitted was from 2 to 38 photoelectrons (see figure \ref{EnergyRecoil1}). The $\chi^{2}$ surfaces are illustrated in figure \ref{FitParabMCvsData} (left) as a function of $R$ and $L_{eff}$ for $\Theta$ = 25$^\mathrm{o}$ (S$_2$ data), 40$^\mathrm{o}$ (S$_2$ data) and 90$^\mathrm{o}$ (TR$_1$ data).  

Following   the method described in \cite{PlanteThesis} a rotated paraboloid is fitted to the  $\chi^2$ map generated around the minimum value for each of the six scattering angles $\Theta_j$. This procedure   to extract the final values for $L_{eff,j}$ and $R_j$  ($j$ = 1...6) averages out small variations on the $\chi^{2}$ surface. The rotated  $\chi^{2}$ is defined as

\begin{eqnarray}
\chi^{2}(L_{eff},R)=\chi^{2}_{min,j}+\bigg[\frac{(L_{eff}-L_{eff,j})\,\cos \omega_{j}-(R-R_{j})\,\sin \omega_{j}}{A_j}\bigg]^{2}\nonumber\\
+\bigg[\frac{(L_{eff}-L_{eff,j})\,\sin \omega_{j}-(R-R_{j})\,\cos \omega_{j}}{B_j}\bigg]^{2}. 
\label{EqRotParaFit}
\end{eqnarray}
where the paraboloid rotation angle $\omega_{j}$ and the parameters $ A_{j}$ and $B_{j}$ are determined  by the fit, together with the minimum $\chi^{2}_{min,j}$.  The location of  $\chi_{min,j}^{2}$ is indicated by the black dots in figure \ref{FitParabMCvsData} (left). The 1$\sigma$  and 2$\sigma$ contours, determined by the values $\chi^{2}_{min}+ 2.3 $ and $\chi^{2}_{min}+6.2$, respectively, are also shown in figure \ref{FitParabMCvsData} (left). 

Figure \ref{FitParabMCvsData} (right) shows the measured nuclear recoil distributions (black dots) and the Monte-Carlo fits (red histograms). The grey zones show the statistical uncertainties on the fit, determined by the 1$\sigma$ contours in figure \ref{FitParabMCvsData} (left). The blue dashed lines correspond to the fit ranges used to compute the $\chi^{2}$. 
The results of the fits are given in table \ref{FitResults} with statistical errors only.

\begin{table}[htb]
\begin{center}
\begin{tabular}{cccccccc}
\hline
$\mathrm{\Theta}$&$L_{eff}$&$\Delta L_{eff} $&$R$&$\Delta R$&$\chi^2$&d.o.f\\
\ [$^{\mathrm{o}}$]&&(stat)&&&&\\
\hline
25&0.386&0.0185&2.26&0.29&18.0&16\\
30&0.305&0.0142&1.84&0.28&19.8&18\\
40&0.285&0.0127&1.86&0.30&20.9&19\\
50&0.294&0.0159&2.79&0.42&36.8&24\\
60&0.283&0.0167&3.43&0.72&33.6&20\\
90&0.301&0.0177&3.49&0.74&15.4&10\\
 \hline
 \end{tabular}
\caption{Results of the  fits. Listed are the optimum values of the scintillation efficiency ${L}_{eff}$ and resolution $R$ as a function of scattering angle $\Theta$. The rms errors are statistical only. The $\chi^2$ values and numbers of degrees of freedom (d.o.f) refer to the least square fit (\protect\ref{eq:chisq}).}
\label{FitResults} 
\end{center}
\end{table}

\subsection{Systematic errors}
\begin{figure}[htb]\begin{center} 
\includegraphics[width=0.35\textwidth]{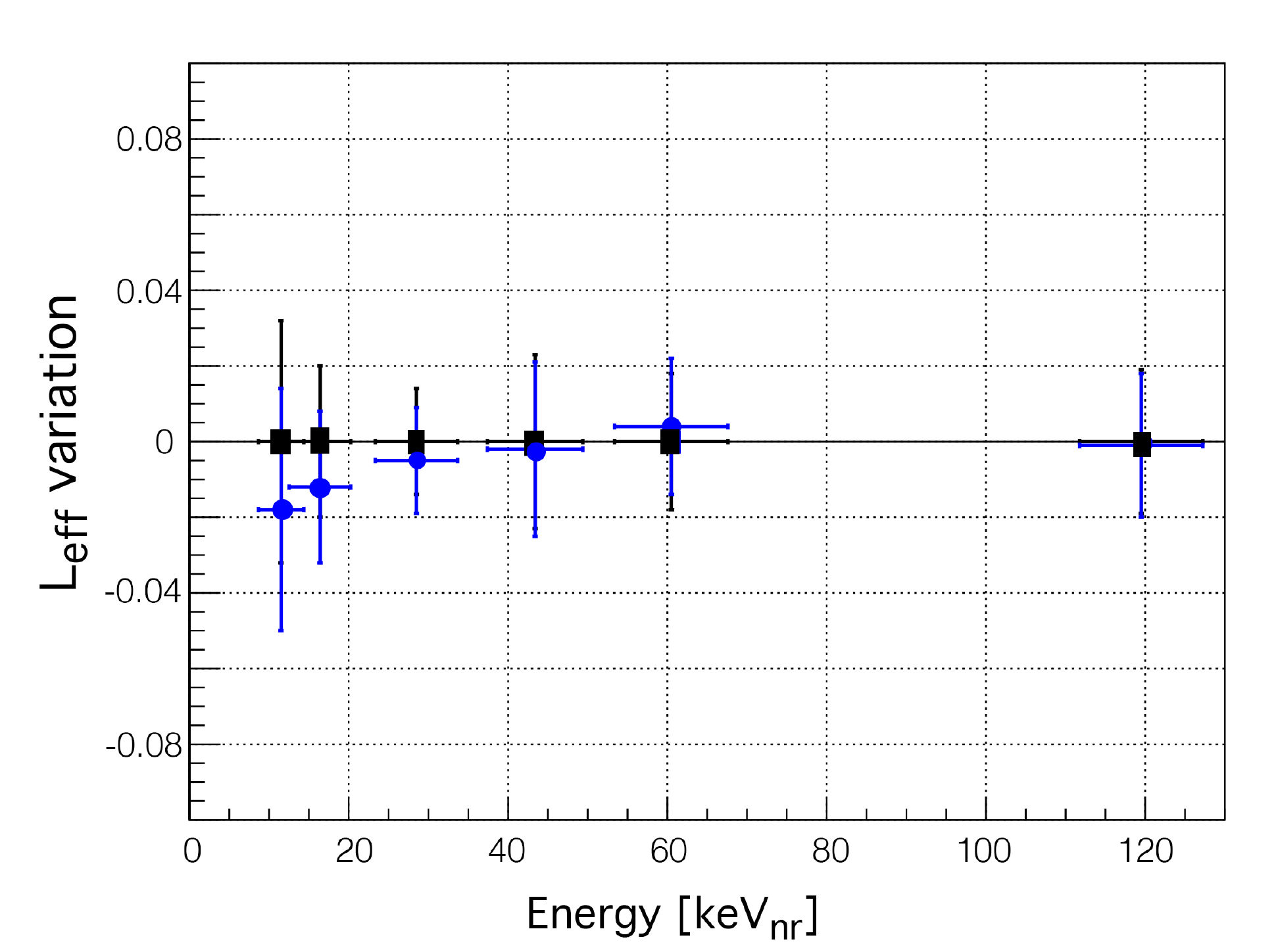}
\caption{\label{LeffVariation} Plot showing how $L_{eff}$ varies when using fixed resolutions $R$  (blue circles) instead of free fit parameters (black rectangles).}
\end{center}\end{figure}

The reliability of the procedure to extract $L_{eff}$ by including the additional fit parameter $R$ has been studied in  \cite{PlanteThesis} for liquid xenon. The analysis  was cross-checked by describing the energy dependence of the resolution with model functions without free parameters. No significant deviation was observed between the values of ${L}_{eff}$ obtained with this method and those determined with the resolution left as a free parameter. 
A  similar test is performed in our work: the measured nuclear recoil spectra with the elastic peak above trigger efficiency roll-off  are fitted with Gaussian functions to obtain the experimental resolution. The simulated contribution of the energy spread due to the finite size of the detector is then subtracted quadratically from the measured resolutions. No significant deviation is observed on $L_{eff}$ between the two methods (figure \ref{LeffVariation}).  

The following sources of systematic uncertainties are taken into account. The uncertainty in the light yield $\Delta Ly$ measured with the $^{241}$Am source  is determined by the error on the calibration constant, estimated to $\pm$2.1\% for the TR$_1$  and $\pm$1.6\% for the TR$_2$ data. The uncertainty $\Delta Trig$ on the trigger efficiency is determined by changing the values of the parameters $b$ and $c$  in (\ref{eq:S}). The error  $\Delta\Theta$ on the angular setting of the LSC is estimated to be  $\pm$ 0.5$\mathrm{^{o}}$. 
The systematic uncertainties are listed in table \ref{TabUncerLeff} together with our final results. The main contributions to the systematic error on $L_{eff}$ stem from the calibration of the americium source and from the trigger efficiency roll-off. The latter large value at 50$\mathrm{^{o}}$ is due to the trigger settings for the TR$_1$ data, for which the effect of the roll-off is more pronounced (see figure \ref{TriggerCurve}). To estimate the systematic error introduced by the uncertainty on the lifetime of the slow component we have varied $\tau_2$ between 1.5 and 1.6 $\mu$s. This leads to an additional (negligible) systematic error of typically 0.003 on $L_{eff}$. The uncertainties in nuclear recoil energies are determined by the angular acceptance of the LSC and are estimated by Monte-Carlo simulation. The total uncertainties on $L_{eff}$ are obtained by  summing  the  statistical errors (table \ref{FitResults}) and systematic errors quadratically.

\begin{table}[htb]
\begin{center}
\begin{tabular}{cc|cc|ccc}
\hline
$\mathrm{\Theta}$[$^{\mathrm{o}}$]&$\mathrm{T_{nr}}$[keV]&$L_{eff}$&$\Delta {L_{eff}}$&$\Delta Ly$&$\Delta Trig$&$\Delta\Theta$\\
\hline
25&11.5$\pm2.8$&0.386&0.032&0.0068&0.0252&0.0028\\
30&16.4$\pm3.9$&0.305&0.020&0.0054&0.0126&0.0010\\
40&28.5$\pm5.2$&0.285&0.014&0.0051&0.0013&0.0007\\
50&43.4$\pm6.0$&0.294&0.023&0.0064&0.0154&0.0006\\
60&60.5$\pm7.1$&0.283&0.018&0.0062&0.0022&0.0004\\
90&119.5$\pm7.7$&0.301&0.019&0.0064&0.0000&0.0000\\
 \hline
 \end{tabular}
\caption{\label{TabUncerLeff} Scintillation efficiency $L_{eff}$ and rms error $\Delta{L}_{eff}$ as a function of scattering angle. The last three columns list the rms systematic errors to $\Delta{L}_{eff}$.}
\end{center}
\end{table}

\subsection{Discussion}
\begin{figure}[htb]\begin{center} 
\includegraphics[width=0.99\textwidth]{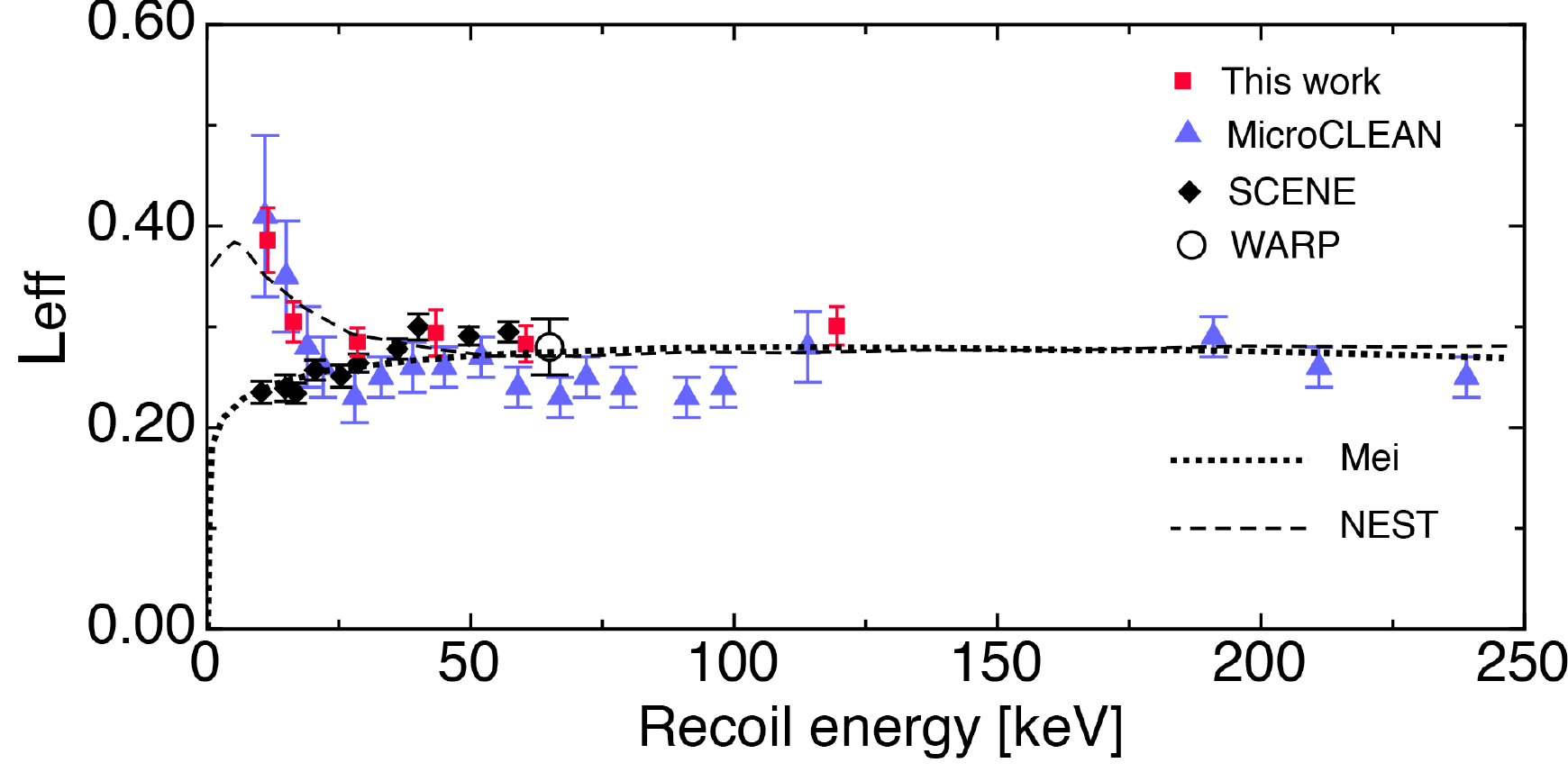}
\caption{\label{ResLeff} Relative scintillation  $L_{eff}$ in liquid argon as a function of nuclear recoil energy from the present experiment (red squares), compared to the previous measurements by MicroCLEAN (blue triangles) \cite{MicroCLEAN}, SCENE (black diamonds) at zero electric field \cite{SCENE} and WARP (open circle) \cite{WARP}. The dotted  and dashed curves show the predictions from the theoretical models by Mei  \cite{MeiPaper} and NEST \cite{NestAr}. }
\end{center}
\end{figure}

The final results for the relative scintillation efficiency from table \ref{TabUncerLeff}  are plotted in figure  \ref{ResLeff}, together with previous measurements. Above $\simeq$25 keV recoil energy $L_{eff}$ is constant with  a mean value $\langle L_{eff}\rangle=0.30\pm0.02$. This result is in fair agreement, although somewhat higher, than that from MicroCLEAN which reports an average value of 0.25 $\pm$ 0.02 above 20 keV \cite{MicroCLEAN}. Our simulations are in excellent agreement with data and we stress here the importance of modelling accurately interactions outside the active volume, in particular for large active volumes (3.14$~\ell$ in \cite{MicroCLEAN} compared to 0.2$~\ell$ in the present work).

Below 20 keV our data show the upturn in $L_{eff}$ already reported by MicroCLEAN  \cite{MicroCLEAN} but not observed by SCENE \cite{SCENE}.  $L_{eff}$ is usually predicted to decrease at low energy due to the combination of energy loss (described by Lindhard's theory) and scintillation quenching (described by Birk's law) \cite{MeiPaper}. This prediction is shown by the dotted curve in figure \ref{ResLeff}. However,  based on the work described in \cite{BezrukovPaper} the NEST group has  recently  applied on argon their model for liquid xenon \cite{Szydagis}. The authors developed  a parametrization to compare with liquid xenon data that reproduce the upturn observed here and  by MicroCLEAN. The upturn could for instance be due to an increasing  exciton-ion ratio at low energies, which would lead to an increasing light yield since it takes less energy to excite  than to ionize \cite{Pcom}. The NEST prediction extrapolated to LAr  \cite{NestAr} is shown by the dashed curve in figure \ref{ResLeff}. 

\section{Conclusions}
Summarizing, the knowledge of  the scintillation efficiency $L_{eff}$ at low nuclear recoil energies is important for direct dark matter searches using noble liquids. We have measured $L_{eff}$ for nuclear recoils, relative to electrons, between  11 keV  and 120 keV  in liquid argon. Single elastic neutron-argon scattering dominates thanks to the small active volume of our argon cell and the well collimated neutron beam, while the contamination from neutron inelastic scattering is negligible. A $\chi^{2}$ minimization is performed leaving $L_{eff}$ and the energy resolution as free parameters. The extraction of $L_{eff}$ is challenging at very low energy where systematic uncertainties on  $L_{eff}$ are determined by the inefficiency of the trigger. At higher energy the uncertainties on  $L_{eff}$ are dominated by statistical errors. The scintillation efficiency is  constant with mean value $\langle L_{eff}\rangle = 0.30\pm0.020$ between 16 keV  and 120 keV. 
The results below 20 keV confirm  the energy upturn reported earlier. The increasing value of $L_{eff}$ will enhance the detection efficiency for low mass WIMPs and be beneficial to dark matter searches using LAr.

\section*{Acknowledgments}
The contribution of P. Otyugova  to the design and installation of the neutron generator facility is gratefully acknowledged. This work was supported by a grant from the Swiss National Science Foundation.

\end{document}